\documentclass[12pt,a4paper]{article}
\usepackage[utf8]{inputenc}
\usepackage[T1]{fontenc}
\usepackage{fix-cm}
\usepackage{amsmath,amssymb}
\usepackage{graphicx}
\usepackage{dcolumn}
\usepackage{bm}
\usepackage{booktabs}
\usepackage[
    colorlinks=true,
    linkcolor=blue,
    citecolor=blue,
    urlcolor=blue,
    filecolor=blue
]{hyperref}
\usepackage{doi}
\usepackage{geometry}
\usepackage{authblk}
\usepackage{float}
\usepackage[numbers,sort&compress]{natbib}
\providecommand{\keywords}[1]{\noindent\textbf{Keywords:} #1}
\providecommand{\pacs}[1]{\noindent\textbf{PACS:} #1}

\newenvironment{acknowledgments}{\section*{Acknowledgments}}{}

\begin{document}

\title{Physics-Informed Neural Networks with Adaptive Kernel Correction for Solving Ill-Posed Fredholm Equations}
\author[1]{Vasily~Yu.~Kozhevnikov\thanks{\texttt{Vasily.Y.Kozhevnikov@ieee.org}}}
\author[1]{Victor~F.~Tarasenko\thanks{\texttt{vf.tarasenko@hcei.ru}}}
\author[1]{Andrey~V.~Kozyrev}
\author[1]{Evgenii~Kh.~Baksht}
\author[2,3]{Vladislav~S.~Igumnov\thanks{\texttt{Vladislav.Igumnov@ieee.org}}}

\affil[1]{Institute of High Current Electronics, Tomsk~634055, Russian Federation}
\affil[2]{Harbin Institute of Technology, Harbin, Heilongjiang~150001, China}
\affil[3]{Tomsk State University of Control Systems and Radioelectronics, Tomsk~634050, Russian Federation}

\date{\today}

\maketitle

\begin{abstract}
In this paper, we propose a robust, mesh-free deep learning framework based on physics-informed neural networks (PINNs) to solve the full inverse problem governed by the Fredholm integral equation of the first kind. The proposed architecture consists of two jointly trained neural networks: one parameterizes the continuous energy spectrum using Fourier features, and the second network adaptively corrects the uncertainty in the empirical integral kernel (e.g., the Tabata-Ito formula for electron-beam attenuation). The method evaluates the Fredholm integral operator directly within a differentiable framework using Gauss-Legendre quadrature, eliminating the need for experimental data interpolation. We rigorously validate the method on synthetic multi-peak spectra, demonstrating high structural stability and accurate peak recovery even under 10~\% random noise and significant data sparsification. We also apply the framework to experimental attenuation data from a nanosecond vacuum diode, revealing complex multi-peak spectral features and anomalous high-energy electrons. The proposed methodology provides a powerful, uncertainty-aware diagnostic tool for indirect charged-particle beam spectroscopy and for solving other ill-posed measurement problems. 
\end{abstract}

\maketitle

\section{Introduction}

One of the widely encountered integral equations of mathematical physics is the Fredholm equation of the first kind \cite{Kulikov_2023}

\begin{equation}
\hat{A} f = \int_{a}^{b} K(x, s) f(s) ds = g(x), \ c\leq x\leq d,
\label{Fredholm}
\end{equation}
which traditionally arises in a broad range of fundamental and applied problems, where the measured (known) function $g(x)$ represents a linear integral transformation of some unknown sought-for function $f(s)$, and the kernel of the integral operator $K(x, s)$ describes some elementary response of the recording system. Classic examples of applications of \eqref{Fredholm} are tomography problems (e.g., positron emission and X-ray computed tomography), where the formation of projection data relative to the spatial distribution of the source is described by Fredholm integral equations in one or two spatial dimensions (the Radon transform) \cite{Hashimoto_2024}. Another application of \eqref{Fredholm} is spectroscopy, where analogous integral equations relate the measured spectrum to a distribution function depending on relaxation times or energy values. Here, reconstructing the unknown distribution function also requires solving the inverse problem for the Fredholm equation \cite{Maradesa_2024}. In addition, solutions of Fredholm equations of the first kind are used to address applied problems in geophysical exploration \cite{Feng_2023}, to enhance and restore blurred images in astronomy \cite{Vasudev_2025} and forensics, and to deconvolve the hardware functions of various measuring instruments. Solving equation \eqref{Fredholm} also arises in reconstructing the radial distribution of the magnetic field amplitude in a Z-pinch from readings of several magnetic sensors located at various distances from the plasma column \cite{Chaikovsky_2005}. Finally, in plasma and beam physics, the Fredholm integral equation of the first kind arises when reconstructing the local spectral energy characteristics of charged particles from integral signals recorded along observed trajectories. A striking example is reconstructing electron energy spectra from attenuation curve data in metal foils of various thicknesses \cite{Baksht_2010, Kozyrev_2015}. 

From the standpoint of physical experiment and measurement theory, the procedure of finding a solution to the Fredholm equation of the first kind represents a nontrivial example of an indirect measurement \cite{Rabinovich_2013}: the directly recorded quantity is the integral response $g(x)$, whereas the sought-for distribution $f(s)$ is not observed directly but can be reconstructed from the relation $\hat{A} f = g$ by inverting the known linear integral operator $\hat{A}$. In contrast to the classical indirect measurement of a scalar quantity, here the measured object is a function, and the inverse problem \eqref{Fredholm} itself is ill-posed; therefore, the reconstruction of $f$ is impossible without additional a priori assumptions and certain regularization procedures, which effectively play the role of a measurement procedure. The result of such a measurement depends not only on the statistics of the directly measured quantities but also on the choice of regularization procedure. Thus, the solution of the Fredholm equation of the first kind \eqref{Fredholm} should be regarded as an indirect measurement of a functional dependence, the validity of which must be confirmed by calibration, direct modeling of the response, and estimation of the uncertainty of the reconstructed distribution $f(s)$. The connection of many indirect physical measurements with the inverse problem for the Fredholm equation of the first kind \eqref{Fredholm} is described in more detail in the classical monograph \cite{Twomey_2019}.

For an abstract operator $\hat{A}: E \to F$, where $E$ and $F$ are metric spaces, according to Hadamard's definition, the problem $\hat{A} f = g$ is called well-posed on the pair of spaces $(E, F)$ if the following three conditions are satisfied: 1) for any $g \in F$ there exists a solution $f \in E$; 2) this solution is unique; 3) the solution depends continuously on $g$; in other words, the convergence $g_n \to g$ in the metric of $F$ implies the convergence of the corresponding solutions $f_n \to f$ in the metric of $E$. If any of these three conditions is violated, the problem is called ill-posed. In other words, the problem $\hat{A}f=g$ is well-posed if and only if the operator $\hat{A}$ has a continuous inverse operator $\hat{A}^{-1}: F \to E$; otherwise, it is ill-posed.

It is important to emphasize that under natural constraints on the kernel $K(x, s)$, for example, in the case of a continuous kernel, the integral operator \(\hat{A}\) is completely continuous (compact) as an operator from the space $C[a, b]$ to the space $C[c, d]$. Since a compact operator on an infinite-dimensional space cannot have a continuous inverse, solving the Fredholm equation of the first kind on this pair of spaces is ill-posed. Moreover, this problem remains ill-posed also on the pair of spaces $L^p(a, b)$ and $L^p(c, d)$. In the applied aspect, the solution of equation \eqref{Fredholm} represents an ill-posed problem precisely because there is no continuous dependence of $f$ on small fluctuations of the right-hand side of the equation, since $g(x)$ represents a directly experimentally measured dependence, which is always known up to the experimental error (systematic and random). An attempt to solve equation \eqref{Fredholm} by fitting the residual of the right-hand side $\left\|\int_{a}^{b} K(x, s) f(s) ds - g(x)\right\|<\delta$ leads to the fact that the resulting distributions $f(s)$ can differ by an arbitrarily large amount. In other words, the fitting methods are unstable with respect to variations in $g(x)$. A similar situation occurs for multidimensional integral equations of the first kind, for example, integrals over closed curves and surfaces; searching for a stable numerical solution to such equations is critical for a broad class of applied problems.

A systematic solution to the ill-posed inverse problem for equation \eqref{Fredholm} is presented in the monograph \cite{Tikhonov_1977}. This approach is commonly referred to as Arsenin–Tikhonov regularization. This approach is based on minimizing the following composite functional: $$M^\alpha[f, g_\delta] = \|\hat{A}f - g_\delta\|^2 + \alpha \Omega[f],$$ where $\hat{A}$ is the exact integral operator (the kernel is assumed to be known without error), \(g_\delta\) is the right-hand side known with some error $\delta$, and $\Omega[f]$ is a stabilizing functional (in the simplest case, the squared norm of the sought-for function), and $\alpha > 0$ is the regularization parameter. The first term in this functional (the residual) ensures that the solution agrees with the experimental data. In contrast, the second term (the stabilizer) ensures the smoothness of the sought-after distribution. It suppresses high-frequency noise, guaranteeing the stability of the result with respect to small changes in the input data. The element $f_\alpha$ at which the minimum of this functional is attained is taken as the regularized solution; here, the parameter $\alpha$ plays the role of a tool for reconciling the accuracy of the data with the required degree of smoothness and is determined, as a rule, by the discrepancy principle from the condition $\|\hat{A}f_\alpha - g_\delta\| < \delta$. If necessary, the stabilizing functional $\Omega[f]$ can be specified in a more general form \cite{Tikhonov_1977} with the inclusion of higher derivatives $\frac{d^k f(s)}{ds^k}$ with nontrivial weight functions $q_k(s)$, which allows additional requirements to be imposed on the sought-for solution $f(s)$

\begin{equation}
M^\alpha[f, g_\delta] = \int_{c}^{d} \left( \int_{a}^{b} K(x, s) f(s) \, ds - g_\delta(x) \right)^2 dx + \alpha \int_{a}^{b} \sum_{k=0}^{n} q_k(s) \left( \frac{d^k f(s)}{ds^k} \right)^2 ds.
\label{Funkzional}
\end{equation}

Obviously, the propagation of uncertainties into equation \eqref{Fredholm} is not limited to the right-hand side alone. A large fraction of the error can come from the integral operator kernel $K(x,s)$ if it is also measured experimentally. According to \cite{Tikhonov_1977}, if the kernel $\hat{A}$ is known only approximately, the original problem $\hat{A} f = g$ is replaced by a problem with an approximate operator $\hat{A}_h$ and an approximate right-hand side $g_\delta$, for which error estimates $\|\hat{A} - \hat{A}_h\| \le h$ and $\|g - g_\delta\| \le \delta$ are specified, and the smoothing functional is constructed using the approximate kernel
$$M^\alpha[f, g_\delta, \hat{A}_h] = \|\hat{A}_h z - u_\delta\|^2 + \alpha \Omega[f],$$
and the regularized solution is taken to be \(f_\alpha\), which minimizes it. Here, the choice of the regularization parameter \(\alpha\) is made differently, since it cannot be determined solely from the right-hand-side residual $\|\hat{A}f_\alpha - g_\delta\| < \delta$. Instead, a generalized discrepancy principle is used, according to which $\alpha$ is chosen so that the residual satisfies $\|\hat{A}_h f_\alpha - g_\delta\| < 2(h\|f_\alpha\| + \delta)$. In effect, this means that the regularization parameter must be consistent with the total error of the right-hand side \(\delta\) and with the kernel error \(h\).

A further mechanism of uncertainty propagation in problem \eqref{Fredholm} arises in the common case where the true integration bounds $[a,b]$ are only approximately known. For instance, if the effective domain is $[a+\Delta a,\, b+\Delta b]$, this perturbation can be interpreted as an additional fluctuation in the kernel of the integral operator associated with the equation. It can be expressed through the difference of integrals $\Delta \hat{A}f = \int_{a}^{a+\Delta a} K(x,s)f(s)ds - \int_{b}^{b+\Delta b} K(x,s)f(s)ds$. It is precisely from this formula that it is evident where the uncertainty penetrates from: it depends not only on the magnitude of the errors \(\Delta a\) and \(\Delta b\) but also on the values of the kernel \(K(x,s)\) and the sought-for function $f(s)$ near the boundaries. This difference is treated as an additional error of the operator $h_{\text{lim}}$, which is majorized by the maximum values of the kernel and the a priori constraints on the solution at the edges of the interval. This error is summed with the error of the kernel itself $h_K$ and the noise of the right-hand side $\delta$, forming the total uncertainty $h_{\text{total}}$, with which the regularization parameter $\alpha$ must also be consistent through the generalized residual $\|\hat{A}_{h_{\text{lim}}} f_\alpha - g_\delta\| \approx 2(h_{\text{total}}\|f_\alpha\| + \delta)$.

From a theoretical point of view, applying the regularization method described in \cite{Tikhonov_1977} yields a solution of \eqref{Fredholm} that is stable with respect to the propagation of all existing uncertainties. The practical application of Arsenin-Tikhonov regularization involves several computational difficulties. By substituting \eqref{Funkzional} into the Euler-Lagrange equation, the following regularization equation is obtained:

\begin{equation}
\begin{aligned}
& \int_{a}^{b} \bar{K}(s, t) f(t) \, dt + \alpha \sum_{k=0}^{n} (-1)^k \frac{d^k}{ds^k} \left( q_k(s) \frac{d^k f(s)}{ds^k} \right) = G(s), \\
& \bar{K}(s, t) = \int_{c}^{d} K(x, s) K(x, t)\, dx,\quad G(s) = \int_{c}^{d} K(x, s) g_\delta(x) \, dx,
\end{aligned}
\label{Tihonov_eq}
\end{equation}
which represents an integro-differential equation with higher-order derivatives of order $n$. Next, the regularization equation \eqref{Tihonov_eq} is solved numerically, requiring the specification of boundary conditions at the ends of the interval $[a, b]$. At this stage, the ambiguity of the interval boundaries $[a, b]$ (see the previous paragraph) is compounded by the need to specify $n-1$ derivative boundary values at both ends of the interval $[a, b]$. These boundary conditions are absent in the original Fredholm equation \eqref{Fredholm}, and it is often impossible to determine their meaning clearly from the conditions of a specific applied problem being solved. The widespread use of Arsenin-Tikhonov regularization methods shows that the final form of the reconstructed solution $f(s)$ depends on the type of boundary conditions in the regularization equation \eqref{Tihonov_eq}.

In most cases, the classical Tikhonov regularization method \cite{Tikhonov_1977} assumes that the resulting regularization equation \eqref{Tihonov_eq} is solved numerically using grid-based finite-difference methods. In real applied problems, the kernel of the integral operator $K(x,s)$ and the experimental data $g(x)$ are discrete sparse sets of experimental points. To apply grid-based numerical methods, the sets of points $K(x,s)$ and $g(x)$ are either approximated by known analytical functional dependencies or artificially densified by interpolation procedures. Such preprocessing introduces substantial additional uncertainties associated with the choice of approximation methods and the heuristics used to estimate values at points where experimental data are absent. Practice shows that such artificial expansion of the sample (for example, turning a sparse set of 10 measurements $g(x)$ into 100–150 points using spline interpolation) radically changes the level of detail of the reconstructed $f(s)$ compared with using a truncated set of a smaller number (20–30) of points. From the standpoint of rigorous measurement theory and metrology for a physical experiment, the artificial expansion (densification) of data sets is also inadmissible, as is the unjustified exclusion of values (thinning of the sample) \cite{Rabinovich_2013}. Any alteration of the experimental data points $g(x)$ introduces uncertainties that cannot be controlled within the classical Arsenin-Tikhonov framework.

In light of the limitations identified above in the classical treatment of the ill-posed Fredholm integral equation of the first kind \eqref{Fredholm}, we now specify the requirements for the method introduced in this study. Specifically, we develop an alternative procedure for computing a regularized solution $f(s)$ to \eqref{Fredholm} that integrates established regularization techniques for ill-posed inverse problems with the representational and optimization capabilities of deep learning. The proposed framework falls within the broader class of physics-informed neural networks (PINNs), whose general methodology is described, for instance, in \cite{Raissi_2019}.

\section{PINN-Based Method Overview}

As noted in the introduction, although the classical Arsenin–Tikhonov regularization framework is grounded in a rigorous mathematical formalism, its application to real-world inverse problems is often constrained by several practical limitations. To overcome these fundamental limitations, the present work proposes a fundamentally different, mesh-free (w.r.t. the experimental $x$-grid) approach based on representing the sought-after spectrum \(f(s)\) as a continuous function parameterized by a deep neural network. The proposed neural-network approach allows one to abandon artificial interpolation and the specification of boundary conditions completely. Also, it provides a natural mechanism to simultaneously account for and compensate for uncertainties in the kernel and the right-hand side of the equation directly during model training. The method proposed below is an improved version of the PINN-based approach previously developed by the authors and tested for reconstructing the spectral energy characteristics of electron beams in a series of works \cite{Kozhevnikov_2026, Kozhevnikov_2026_preprint1, Kozhevnikov_2026_preprint2, Tarasenko_2026}. 

The central element of the proposed approach is abandoning the discrete representation of the sought-after function in favor of its continuous parameterization via a deep neural network. We shall seek the spectrum \(f(s)\) in the form of a function \(f_\theta(s)\), where \(\theta\) is the set of trainable weights and biases of the deep neural network. This mesh-free approach automatically avoids the need to specify boundary conditions for the regularization equation. To improve the approximation accuracy and combat the well-known spectral bias effect, which manifests itself in the tendency of neural networks to smooth out high-frequency details, the network input is not the coordinate \(s\) itself but its nonlinear transformation (Fourier features). The input vector is formed by projecting \(s\) onto a set of sine and cosine functions with various frequencies:
$$
 \gamma(s) = \left[ \sin(2\pi s \mathbf{B}), \cos(2\pi s \mathbf{B}) \right], 
$$
where the matrix \(\mathbf{B}\) consists of randomly initialized frequencies fixed for the duration of training. This allows the network to efficiently capture both the macroscopic structure of the spectrum and its fine details, which is critical for reconstructing real physical distributions.

In addition, since by its physical meaning the spectral density of particles cannot be negative, a Softplus activation function is applied at the network output, guaranteeing the satisfaction of the condition \(f_\theta(s) \ge 0\) for all \(s \in [a, b]\). This a priori assumption is necessary for obtaining physically meaningful regularized solutions to a Fredholm equation of the first kind, so the final representation of the sought-after function takes the form:
$$
 f_\theta(s) = \text{Softplus} \left( \text{MLP}\big(\gamma(s)\big) \right), 
$$
where $\text{MLP}(\cdot)$ denotes a multilayer perceptron. 

The second key element of the proposed method is the natural accounting for the integral operator kernel uncertainty \(K(x,s)\). In real physical experiments, both $g(x)$ and $K(x,s)$ are known only to some accuracy, which the following relations can express:
\begin{equation}
\begin{gathered}
g(x) = \bar{g}(x) \pm \Delta g(x), \\
K(x, s) = \overline{\overline{K}}(x, s) \pm \Delta K(x, s),
\end{gathered}
\label{errors}
\end{equation}
where \(\bar{g}(x)\) and \(\overline{\overline{K}}(x,s)\) are the mean values of the right-hand side of the equation and the kernel, and \(\Delta g(x)\) and \(\Delta K(x,s)\) are the corresponding uncertainty quantities (noise or experimental error). In classical theory \cite{Tikhonov_1977}, accounting for both uncertainties requires introducing integral error estimates \(\delta\) and \(h\) and using a generalized discrepancy principle. Instead of specifying these estimates a priori and then attempting to reconcile the regularization parameter with them, we parameterize the correction to the kernel \(\Delta K_\phi(x, s)\) by means of a second deep neural network that takes the coordinates \(x\) and \(s\) as input. Hereafter, the network $f_\theta$ is referred to as the spectrum PINN, and the network \(\Delta K_\phi\) as the auxiliary kernel-correction network; only the former is physics-informed in the strict sense, whereas the latter implements a trainable, physically constrained correction to the empirical kernel. The final kernel of the integral operator is formed as the sum of the measured function \(K(x,s)\) and the trainable correction
$$
 K_{\text{total}}(x, s) = K(x, s) + \Delta K_\phi(x, s), 
$$
which allows the algorithm to adapt the kernel during training within the limits specified by the experimental error level \(\Delta K(x,s)\). The regularization of the magnitude of the correction \(\Delta K_\phi\) is controlled by a separate term in the generalized loss function, which keeps the solution close to \(\overline{\overline{K}}(x,s)\) and allows the error to be compensated. Thus, the proposed approach implements the Arsenin-Tikhonov idea of accounting for the inaccuracy of the operator \(\hat{A}_h\) in a more flexible form. Instead of rigidly choosing the parameter \(h\) and computing \(hR + \delta\), we let the neural network correct the kernel.

Since the sought-after function \(f_\theta(s)\) is specified continuously everywhere on the interval \([a, b]\), we can compute the integral of it with high accuracy at any point \(x\). For this purpose, Gauss-Legendre quadrature with a fixed number of nodes \(N_q\) on the interval \([a, b]\) is used. The approximate value of the measured response at the \(i\)-th experimental point \(x_i\) is computed as:
$$
 g_{\text{pred}}(x_i) = \sum_{j=1}^{N_q} w_j K_{\text{total}}(x_i, s_j) f_\theta(s_j), 
$$
where \(s_j\) and \(w_j\) are the nodes and weights of the Gauss-Legendre quadrature, respectively. Thus, we integrate the continuous function over the entire specified interval and compare the result only with those values \(g_\delta(x_i)\) that were actually measured in the experiment. We do not need to construct a regular grid in \(x\) and artificially densify the experimental data — this eliminates errors associated with the heuristics of the points of the original sparse sets \(g(x)\).

The training of the neural networks is carried out by minimizing the generalized loss function, which represents an adaptation of the Tikhonov smoothing functional \eqref{Funkzional} to the deep learning process:
\begin{equation}
 L = L_{\text{data}} + \lambda_1 L_{\text{reg\_f}} + \lambda_2 L_{\text{reg\_K}}, 
\label{Losses}
\end{equation}
where \(\lambda_1\) and \(\lambda_2\) are positive hyperparameters that serve as analogs of the regularization parameters $\alpha$. The first term \(L_{\text{data}}\) represents the normalized residual of the right-hand side of equation \eqref{Fredholm} \(\|\hat{A}f - g_\delta\|^2\) in functional \eqref{Funkzional}:
$$
 L_{\text{data}} = \frac{\frac{1}{N_{\text{exp}}} \sum_{i=1}^{N_{\text{exp}}} \left( g_{\text{pred}}(x_i) - g_\delta(x_i) \right)^2}{\frac{1}{N_{\text{exp}}} \sum_{i=1}^{N_{\text{exp}}} g_\delta(x_i)^2 + \varepsilon}, 
$$
where \(N_{\text{exp}}\) is the number of experimental points $g(x)$, and \(\varepsilon\) is a small constant ensuring numerical stability when working with data $g(x)$ close to zero. Normalizing by the mean square of the experimental data makes the loss function scale-invariant, which accelerates and stabilizes training. 

The second term \(L_{\text{reg\_f}}\) is a discrete analog of the stabilizing functional \(\Omega[f]\). It penalizes both large values of the function itself and its sharp changes (high-frequency noise):
$$
 L_{\text{reg\_f}} = \frac{q_0 \, \mathbb{E}\left[ f_\theta(s)^2 \right] + q_1 \, \mathbb{E}\left[ \left( \frac{df_\theta(s)}{ds} \right)^2 \right]}{\mathbb{E}\left[ g_\delta(x)^2 \right] + \varepsilon}, 
$$
where \(\mathbb{E}[\cdot]\) denotes the sample mean over the quadrature nodes (for an arbitrary function \(h(s)\), \(\mathbb{E}[h(s)] = \frac{1}{N_q} \sum_{j=1}^{N_q} h(s_j)\)). This form of the stabilizer corresponds to choosing weight functions \(q_0 = 1\) and \(q_1 = 0.1\) in the general expression \eqref{Funkzional}. The choice \(q_0 = 1\) sets a natural scale for normalizing the sought-after function $f(s)$. The weight coefficient \(q_1 = 0.1\) for the derivative is chosen because too strong a penalty on the derivative (\(q_1 \geq q_0\)) would lead to excessive smoothing and blurring of the characteristic features of $f(s)$. In contrast, a moderate value \(q_1 = 0.1\) suppresses potential high-frequency artifacts while preserving the fine structure of the distribution. Normalization by the mean square of the experimental data \(\mathbb{E}\left[ g_\delta(x)^2 \right]\) in the denominator of the stabilizer guarantees that both terms (both the norm of the function and its derivative) make a real contribution to the penalty without canceling each other, which makes \(L_{\text{reg\_f}}\) a direct discrete analog of the stabilizer \(\Omega[f]\).

Finally, the third term \(L_{\text{reg\_K}} = \mathbb{E}\left[ \Delta K_\phi(x, s)^2 \right]\) limits the magnitude of the correction to the kernel, implementing the idea of the generalized discrepancy principle with allowance for errors in the kernel of the integral operator. This term limits the degree of freedom in modifying the kernel and keeps the final kernel \(K_{\text{total}}\) close to \(K(x,s)\). 

All computations and joint training of both neural-network models were carried out using the NVIDIA PhysicsNeMo software platform (PyTorch backend). High-performance computations were performed on a 16 GB NVIDIA RTX A4000 graphics accelerator. The \textit{SpectrumNet} architecture (the spectrum PINN) consists of 4 hidden layers with 128 neurons each (\texttt{tanh} activation) with a preliminary Fourier transformation (32 frequencies). The \textit{DeltaKNet} architecture (the auxiliary kernel-correction network) consists of 4 hidden layers with 256 neurons each (\textit{tanh} activation); the output layer is linear (without activation) to allow the correction $\Delta K$ to take both positive and negative values. To minimize the loss function, we used the adaptive Adam optimizer with an initial learning rate of $10^{-3}$. The learning rate was reduced stepwise (with a factor of $\gamma = 0.999$) at fixed intervals (every 100 epochs). An early-stopping mechanism controlled the training process: if the loss did not improve for 2000 epochs, training was terminated, and the model with the lowest error on the validation set was saved. A randomly selected portion of the $g(x)$ experimental points (10–20 \%) that did not participate in training was used as the validation set. For rigorous verification of the adequacy of the algorithm, training and validation were carried out both on synthetic data sets with an a priori known model solution \(f(s)\) and on real experimental data sets for which the classical Arsenin-Tikhonov regularization technique \cite{Tikhonov_1977} was successfully used, for example, the results of works \cite{Kozyrev_2015, Baksht_2010}.

\section{Demonstration of the Method’s Capabilities Using Synthetic Model Data}

Studying the energy characteristics of subnanosecond electron beams generated in vacuum and gas-filled pulsed diodes under high-voltage pulses with a steep leading edge is a topical problem in modern plasma and beam physics. Of particular interest is the generation of electrons with anomalously high energies ("anomalous" electrons) $s > qU_0$ ($q$ is the electron charge, $U_0$ is the maximum voltage amplitude across the gap). The possibility of fractions of such electrons occurring in fast discharges has been substantiated theoretically \cite{Kozhevnikov_2022, Kozyrev_2018}. Since direct measurement of the spectral characteristics of electron beams of short nanosecond and subnanosecond durations is associated with serious technical difficulties, the main method of their diagnostics has historically remained the attenuation curve method, based on recording the charge or current of the beam that has passed through metal foils of various thicknesses \cite{Baksht_2010, Kozyrev_2015}. The scheme of the attenuation curve method is shown below in Fig.~\ref{fig:attenuation_curves_method}

\begin{figure}[!htb]
\centering
\includegraphics[scale=.45]{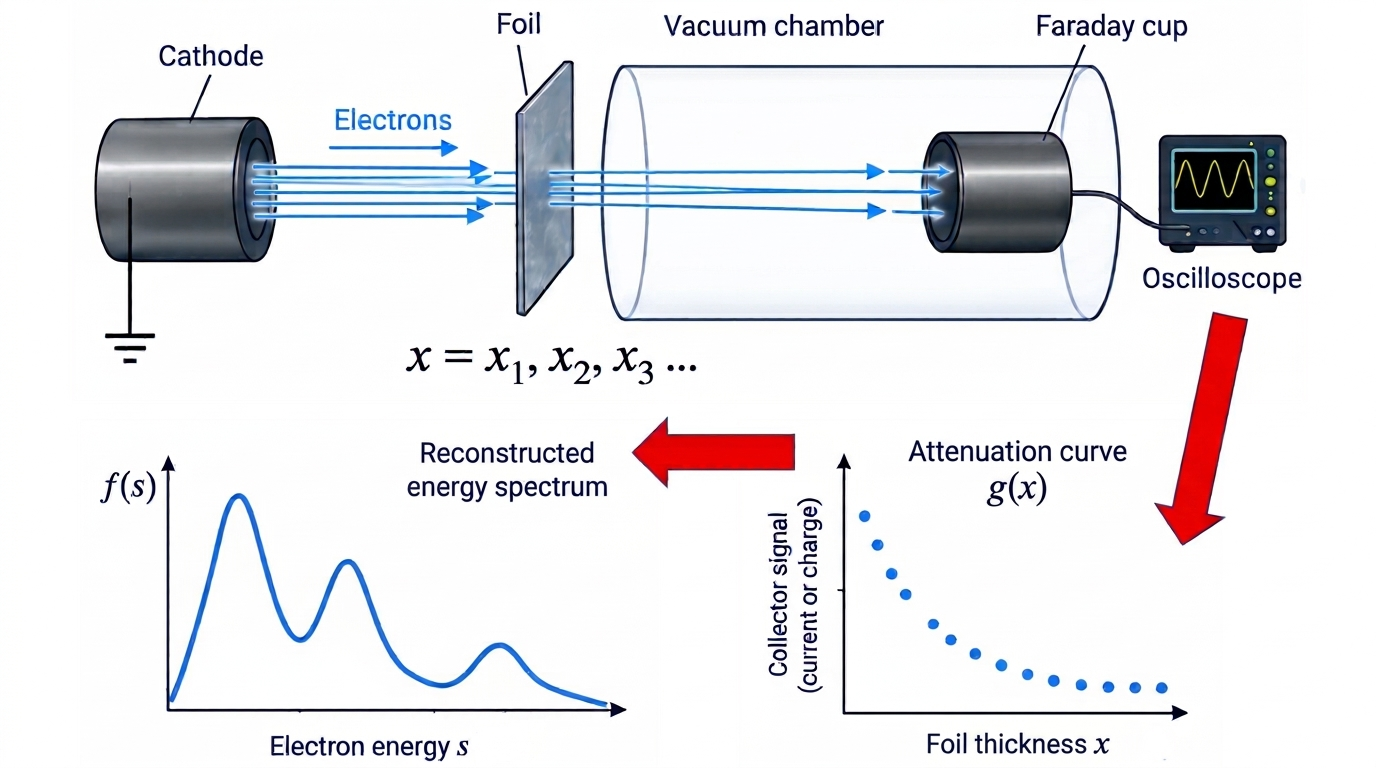}
\caption{\label{fig:attenuation_curves_method} Schematic representation of the indirect measurement process of the spectral energy characteristics of electron beams (attenuation curve method).}
\end{figure}

The reconstruction of the sought-after energy spectrum $f(s)$ from these data represents a classical ill-posed inverse problem for the Fredholm equation of the first kind \eqref{Fredholm}, the kernel of which is the attenuation function of monoenergetic electron beams known as the empirical Tabata-Ito formula \cite{Tabata_1975}

\begin{equation}
\begin{gathered}
K(x, s) = \frac{1 + e^{-E(s)}}{1 + \exp\left[ \big(E(s) + 2\big) \dfrac{x}{\mu(s)} - E(s) \right]}, \\
E(s) = 5.863 \exp\left(-\frac{0.0671}{1 + 0.146 s^{1.86}}\right), \quad \mu(s) = 2.46 s \left(1 - \frac{0.9878}{1 + 3.83 s}\right),
\end{gathered}
\label{kernel}
\end{equation}
where the electron energy \(s\) is in MeV, and the foil thickness \(x\) is in centimeters. 

The energy spectra of electrons in fast-occurring electrical discharges have a nontrivial multi-peak structure, whose identification is complicated by experimental noise and errors. Before applying the algorithm proposed above to real experimental data, we carried out dozens of cycles of thorough validation on controlled synthetic sets of model spectra $f_{\text{model}}(s)$. Below we illustrate the validation of the method on a synthesized artificial data set: we specify an a priori known model spectrum \(f_{\text{model}}(s)\), then compute the right-hand side of equation \eqref{Fredholm} with its help, and then reconstruct this spectrum using the proposed PINN-based algorithm.

The Tabata-Ito formula approximates experimental data sets and has its own error. According to the original paper \cite{Tabata_1975}, the root-mean-square absolute error of this equation is 0.03 in most cases. For aluminum in the energy range from 5~keV to 500~keV, the maximum relative error of the Tabata-Ito formula can reach 10–15~\%. This level of uncertainty must be incorporated into the model to reconstruct the spectral energy characteristics. 

The following analytical function $f_{\text{model}}(s)$ was chosen as the model distribution, which imitates a complex three-peak structure in the energy range $s \in [5, 500]$~keV
\begin{equation}
f_{\text{model}}(s) = \frac{50 {\mathrm e}^{-\frac{25 \left(10 s-1\right)^{2}}{2}}+100 {\mathrm e}^{-50 \left(5 s-1\right)^{2}}+70 {\mathrm e}^{-\frac{25 \left(20 s-7\right)^{2}}{32}}}{100+57 {\mathrm e}^{-\frac{25}{2}}}.    
\label{f_model}
\end{equation}

After specifying the analytical form of the model spectrum $f_{\text{model}}(s)$, we compute a synthetic attenuation curve. For this purpose, the function $f_{\text{model}}(s)$ is numerically integrated with the Tabata-Ito kernel in accordance with equation \eqref{Fredholm}: 
\begin{equation}
g_{\text{model}}(x_i) = \int_{a}^{b} K(x_i, s) f_{\text{model}}(s) \, ds,
\label{g_model}
\end{equation}
where the integration is performed using Gauss-Legendre quadrature. The set of foil thicknesses \(x_i\) for which the right-hand side is computed imitates a real nonuniform experimental sample. In this example, to test the method, a nonuniform distribution of 21 points in order of increasing foil thickness was chosen. The set of points of the attenuation curve \eqref{g_model} obtained in this way is fed into the PINN-based spectrum reconstruction algorithm.

\begin{figure}[!htbp]
\vspace*{-1.5em}
\centering
\includegraphics[width=0.41\textwidth]{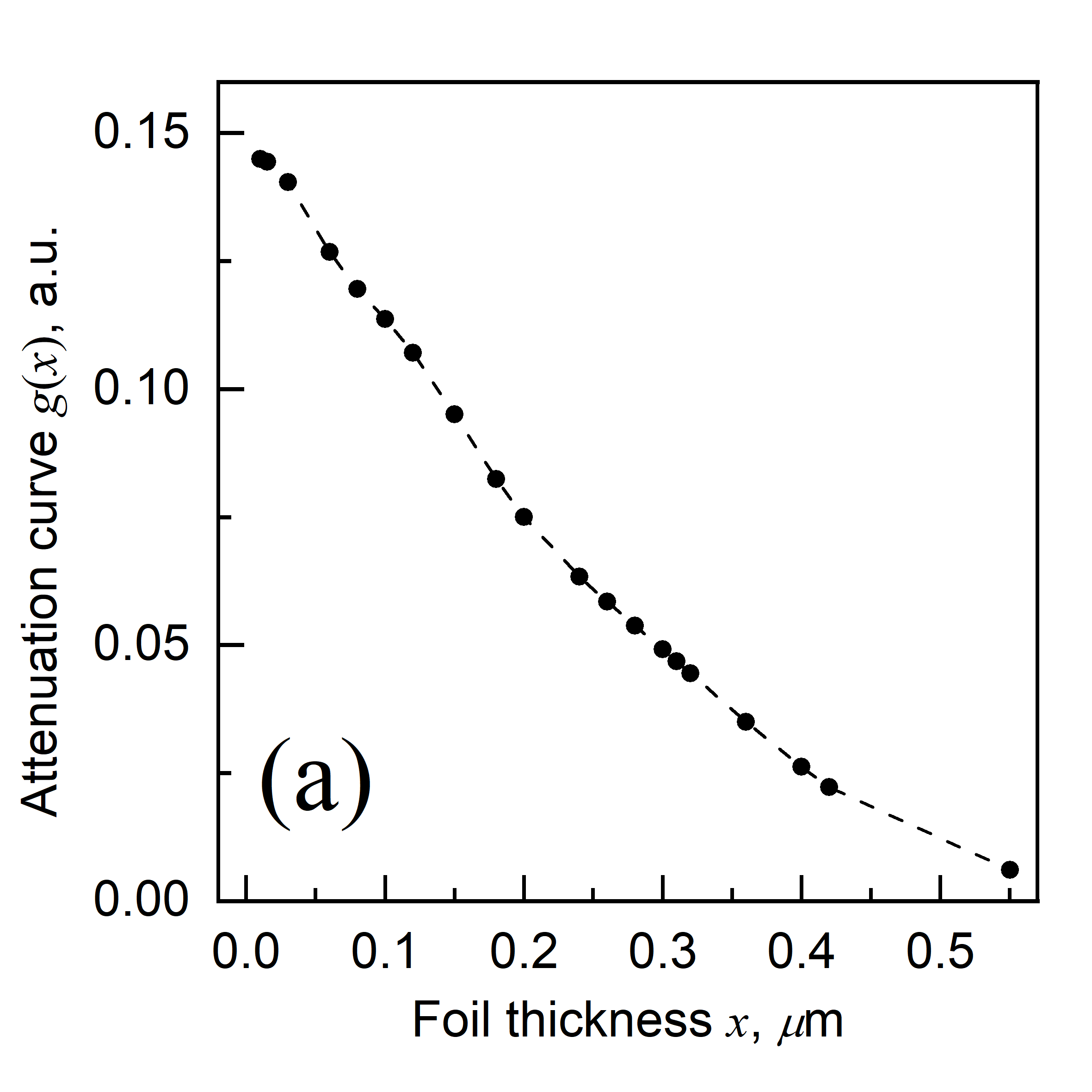}\hspace{0.02\textwidth}%
\includegraphics[width=0.41\textwidth]{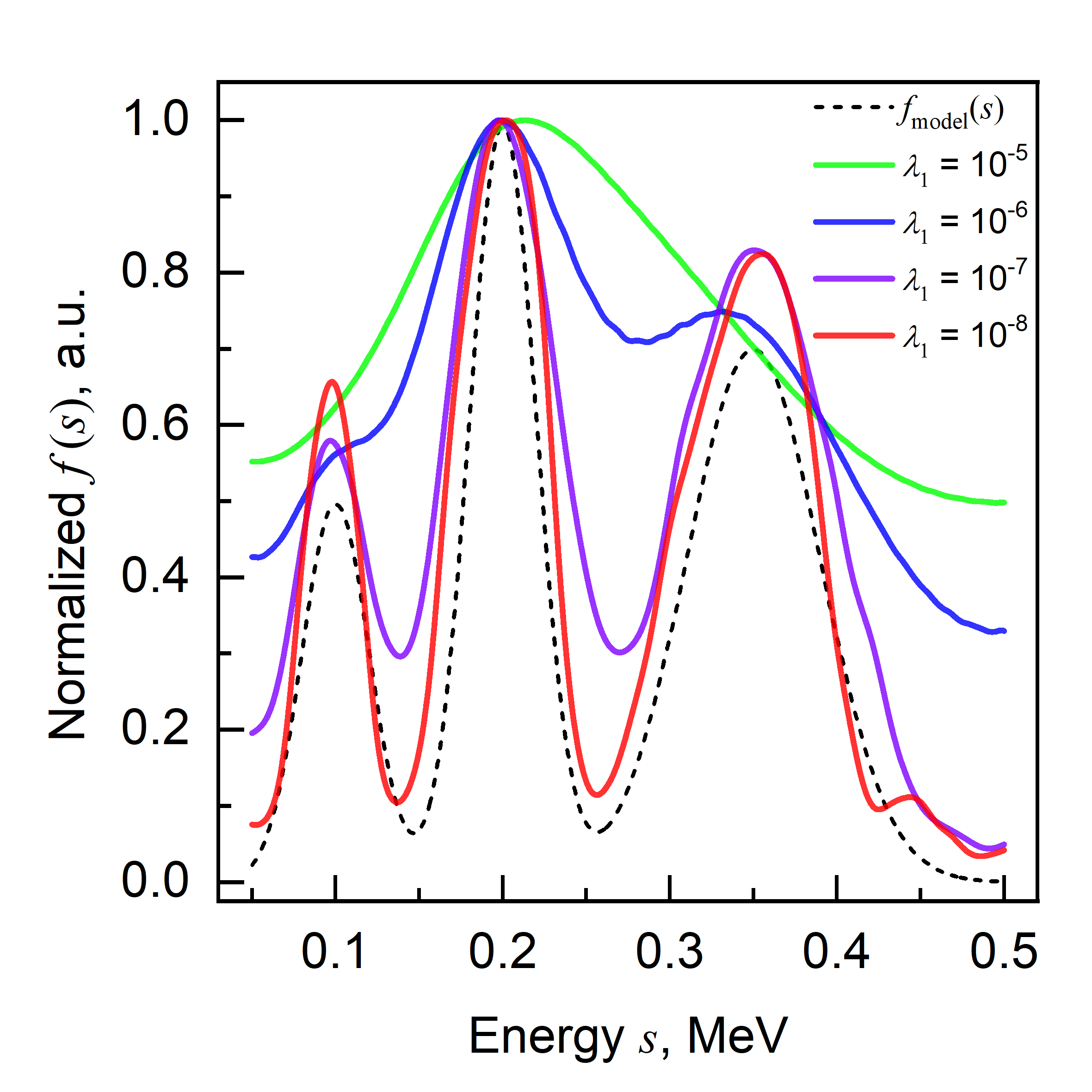}\\[-4ex]

\includegraphics[width=0.41\textwidth]{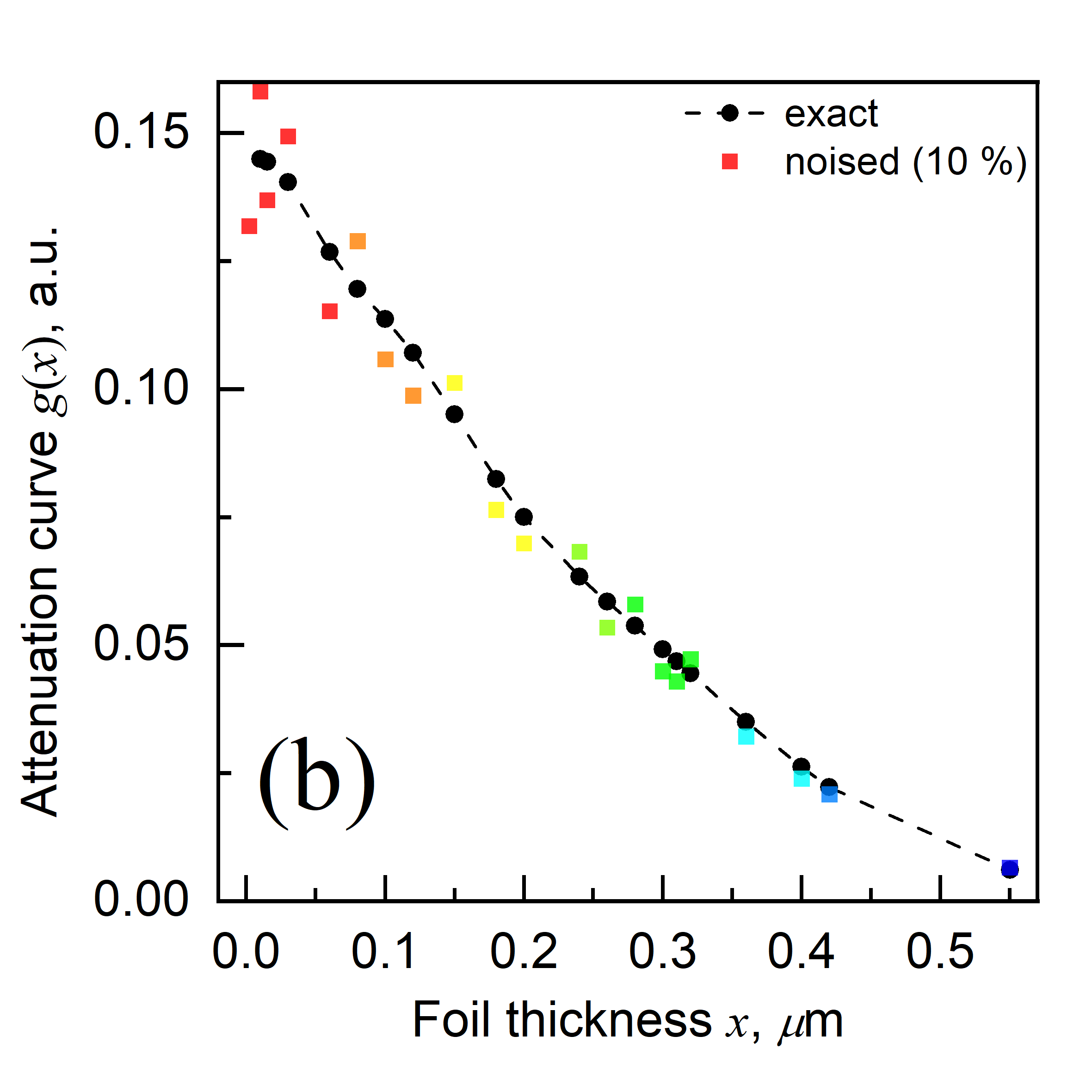}\hspace{0.02\textwidth}%
\includegraphics[width=0.41\textwidth]{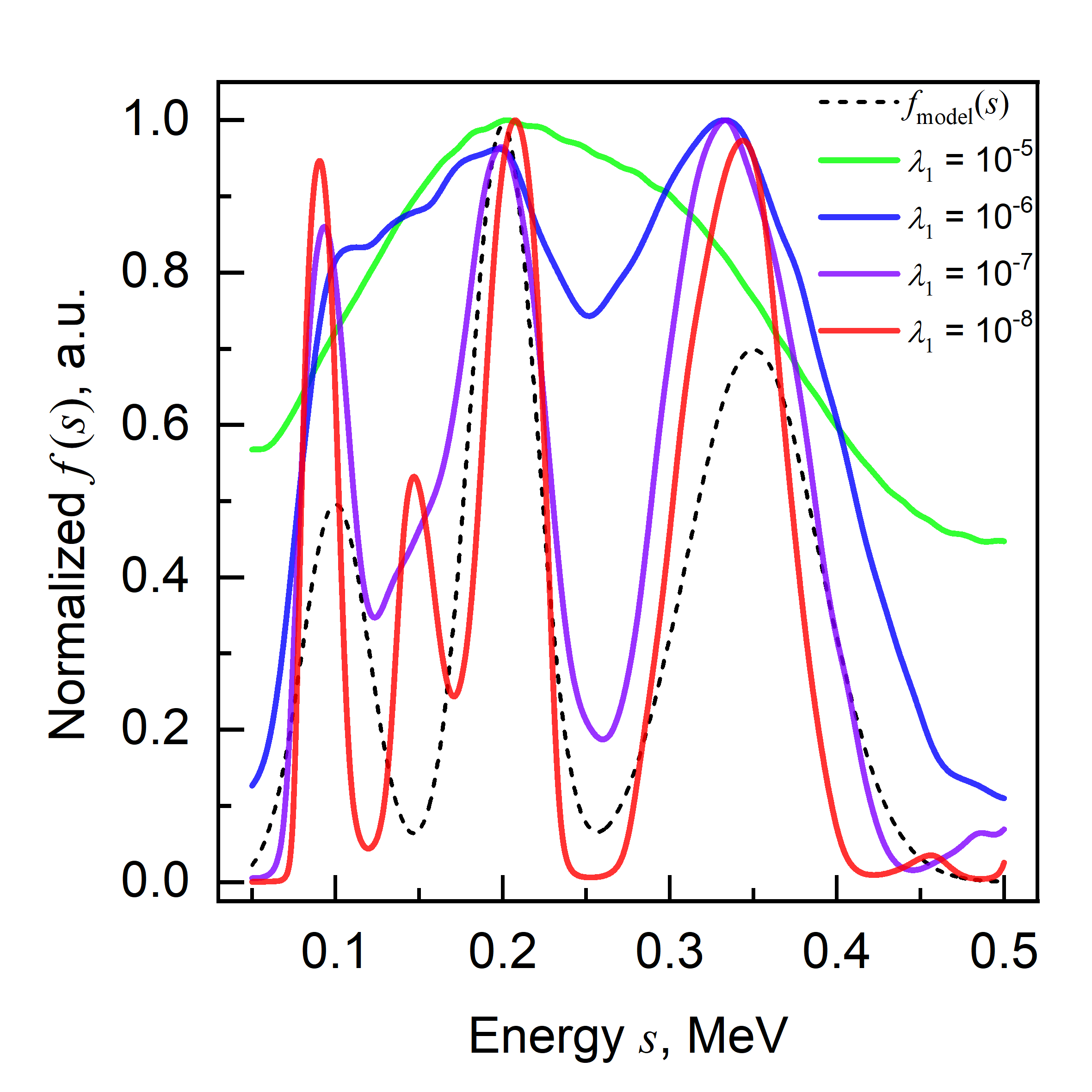}\\[-4ex]

\includegraphics[width=0.41\textwidth]{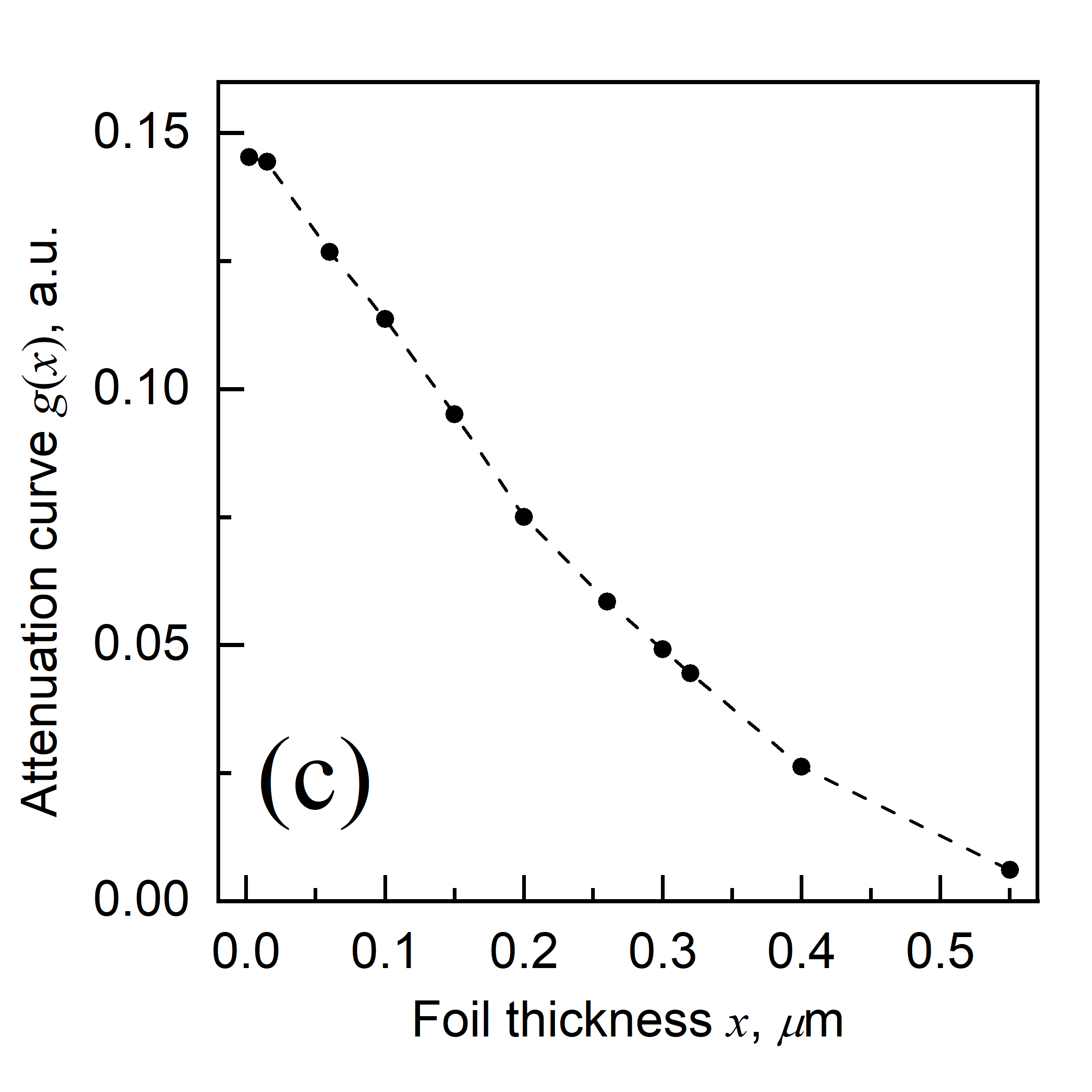}\hspace{0.02\textwidth}%
\includegraphics[width=0.41\textwidth]{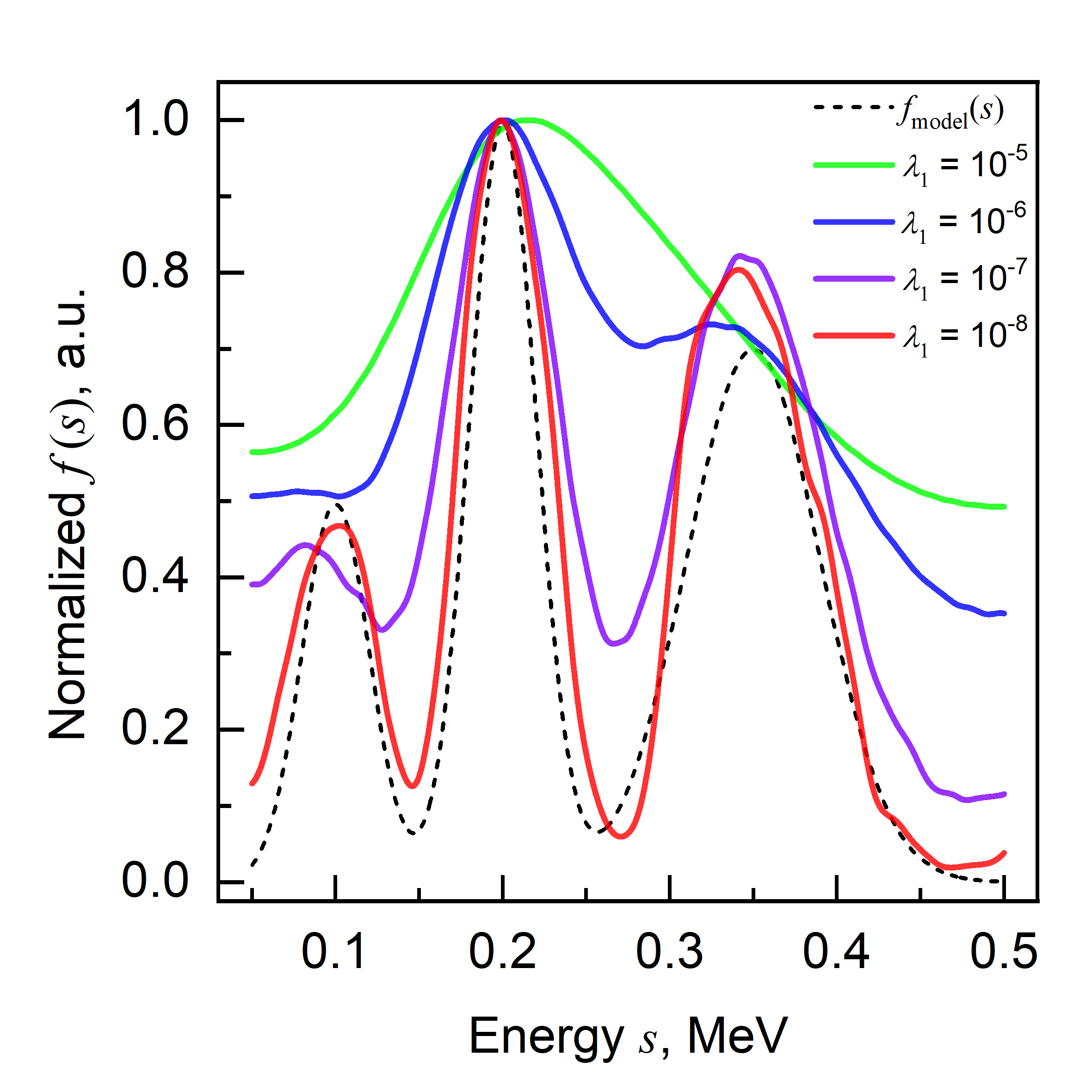}\\[-4ex]

\includegraphics[width=0.41\textwidth]{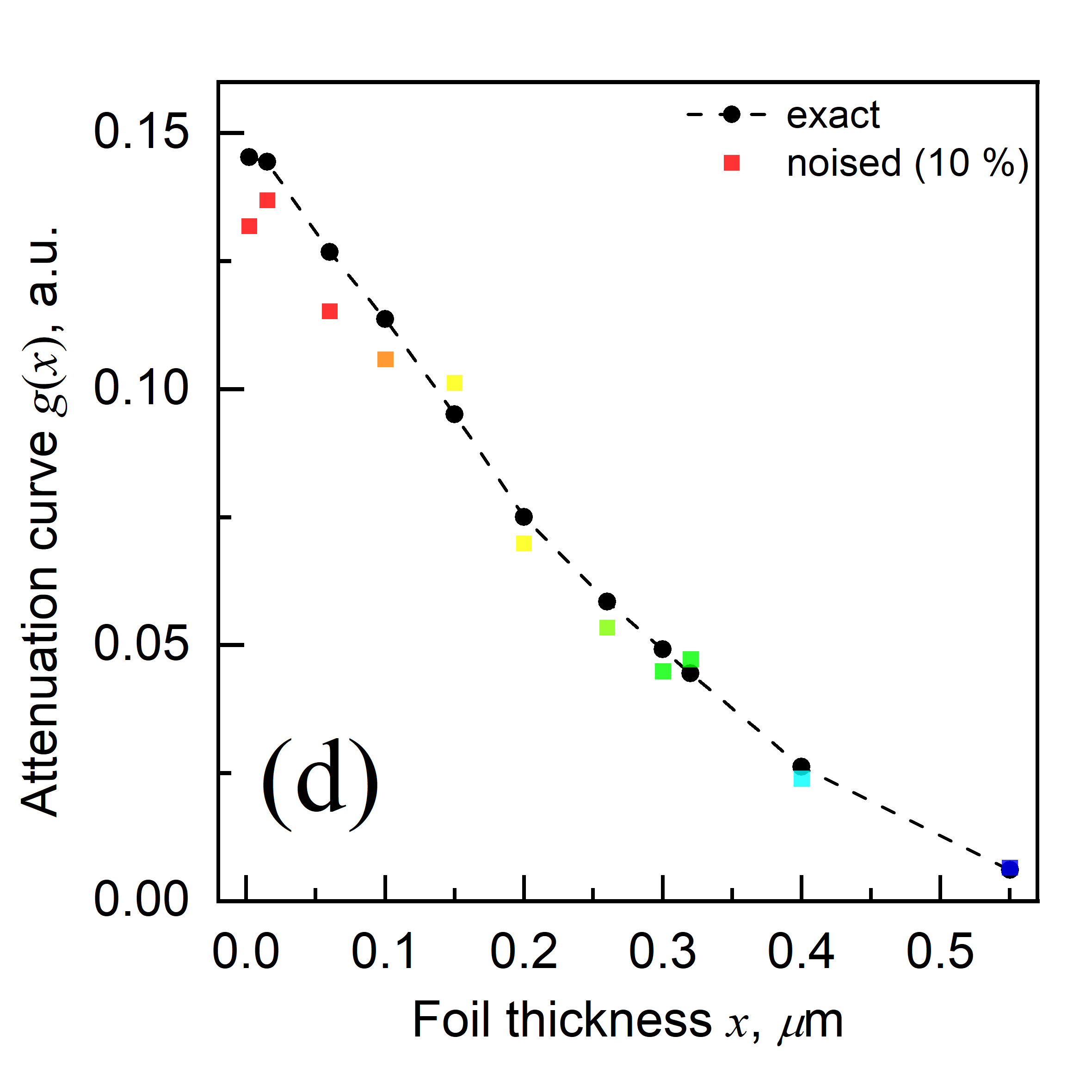}\hspace{0.02\textwidth}%
\includegraphics[width=0.41\textwidth]{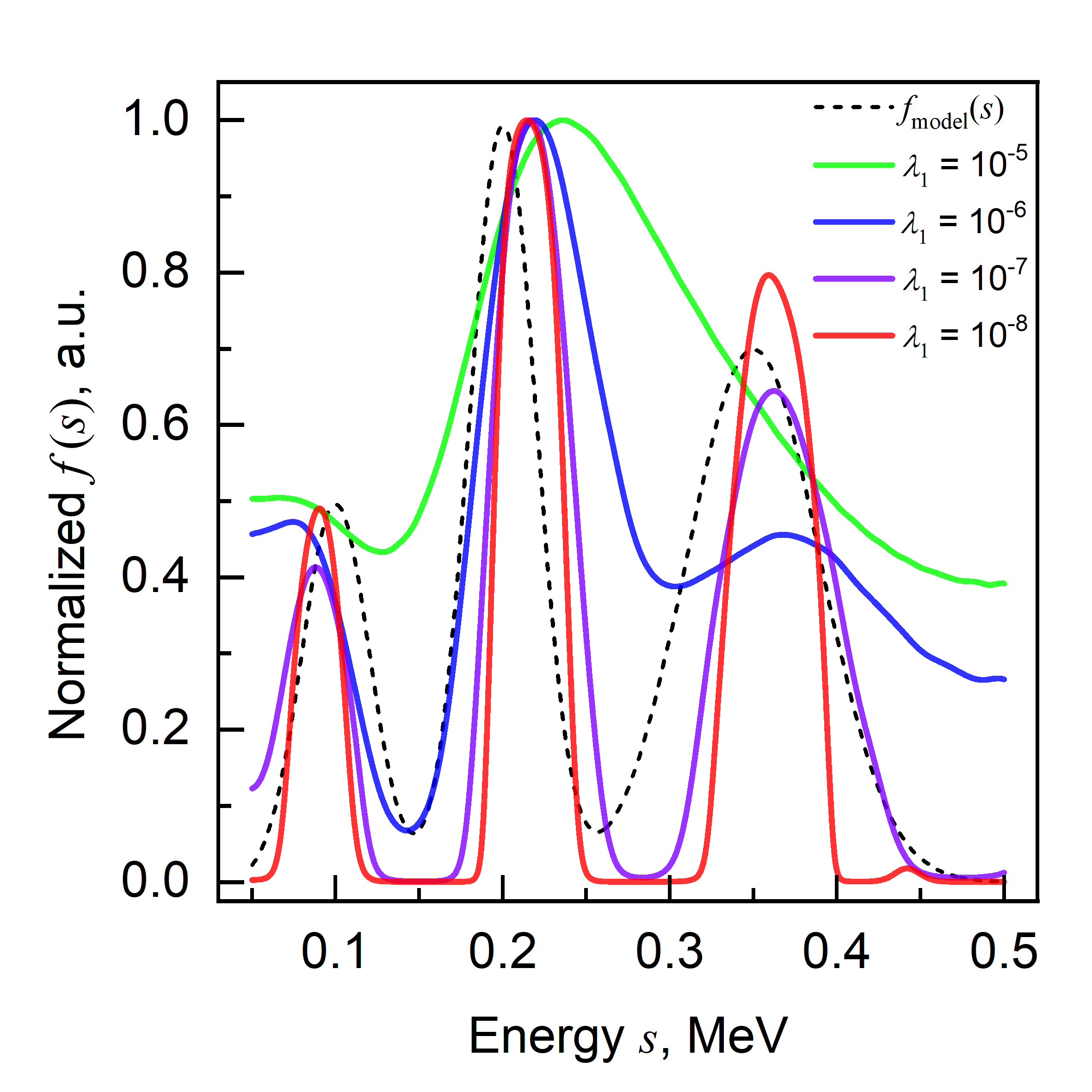}\\[-1ex]

\caption{\label{fig:validation} Validation on synthetic data: attenuation curves \(g(x)\) (left column) and reconstructed spectra \(f(s)\) for several \(\lambda_1\) (right column) for (a) original data, (b) 10\,\% noise, (c) half the number of points, and (d) both factors; dashed line — model spectrum \(f_{\text{model}}(s)\).}
\end{figure}

Fig.~\ref{fig:validation} presents the results of validating the spectrum reconstruction method (right column) for four modifications of the original attenuation-curve data set (left column). The reference case (a) represents the reconstruction of the spectrum from 21 points of the chosen attenuation curve \eqref{g_model}, which is calculated from the model spectrum function \eqref{f_model}. Next, the effect of introducing an error in the form of 10~\% random noise into the amplitudes of the attenuation curves is shown (case (b)); the effect of halving the number of points of the reference set (a) is shown in case (c); and the effect of both of these factors on the reconstructed spectrum is shown in case (d).

In each of the listed cases, the spectrum reconstruction by the proposed PINN-based approach was performed for four different values of the regularization parameter $\lambda_1\ = 10^{-5}$, $10^{-6}$, $10^{-7}$, $10^{-8}$ to demonstrate the gradual increase in reconstruction accuracy. At \(\lambda_1 = 10^{-5}\) (green curve) and even larger values, the main maximum appears in the spectrum against the general smoothed background, its position corresponding to that in the model distribution. Then, upon passing to \(\lambda_1 = 10^{-6}\) (blue curve), the outlines of the side peaks at 0.1~MeV and 0.35~MeV begin to manifest themselves in the reconstructed spectrum. The best agreement with the model spectrum is demonstrated by the curves at \(\lambda_1 = 10^{-7}\) and below: all three maxima are clearly manifested on them, their positions exactly coinciding with the positions of the maxima of the model spectrum \ref{f_model} — approximately \(0.1\), \(0.2\), and \(0.35\) MeV. The normalized spectrum values at the boundaries of the energy interval are also reproduced correctly. Even with 10~\% error in the initial data, the method reliably identifies the beam's energy composition (the presence and position of the peaks). However, quantitative amplitude estimation requires optimal selection of the regularization parameter based on the actual noise level. Naturally, reconstructing spectra from noisy data (c) introduces noise features that distinguish the obtained target spectrum from the model shape. Still, the most important advantage is the ability to determine the positions of the maxima of all spectral energy groups. 

To quantitatively assess the accuracy of the spectrum reconstruction in each scenario, we used two metrics computed at the points of the experimental attenuation curve: the relative error in the \(L_2\) norm and the maximum relative deviation. Table \ref{tab:lambda1_sweep} presents the dependence of these metrics on the regularization parameter \(\lambda_1\) for the proposed method. Here, we fixed \(\lambda_2 = 0.01\) for all runs. This hyperparameter plays the role of a penalty for the deviation of the adaptive correction of the equation kernel \(\Delta K_\phi(x,s)\) from the mean value. Empirically, the value of the hyperparameter \(\lambda_2 = 0.01\) corresponds to the admissible level of relative error of the kernel \(\sim 10\)~\%, characteristic of the values of the Tabata-Ito formula \cite{Tabata_1975} for aluminum ($Z = 13$), which limits the adjustment of the kernel shape to the noise of the right-hand side of the equation. 

\begin{table}[!htb]
\centering
\begin{tabular}{l c c c c}
\toprule
\(\lambda_1\) & \textbf{(a)} & \textbf{(b)} & \textbf{(c)} & \textbf{(d)} \\
\midrule
\(10^{-5}\) & 2.10 / 3.16 & 13.43 / 17.36 & 1.68 / 2.48 & 5.73 / 5.90 \\
\(10^{-6}\) & 1.14 / 1.46 & 13.05 / 17.36 & 0.99 / 1.06 & 4.98 / 5.03 \\
\(10^{-7}\) & 0.57 / 0.73 & 12.72 / 17.36 & 0.49 / 0.55 & 4.40 / 4.15 \\
\(10^{-8}\) & 0.21 / 0.31 & 12.48 / 17.36 & 0.15 / 0.20 & 4.20 / 3.87 \\
\(10^{-9}\) & 0.12 / 0.17 & 12.42 / 17.36 & 0.07 / 0.08 & 4.15 / 3.87 \\
\bottomrule
\end{tabular}
\caption{Dependence of the spectrum reconstruction accuracy on the regularization parameter \(\lambda_1\) for the proposed method with a fixed kernel representation accuracy parameter \(\lambda_2 = 0.01\). The values are given in the format: relative \(L_2\) error (\%) / maximum relative deviation \(\delta_{\max}\) (\%).}
\label{tab:lambda1_sweep}
\end{table}

The data in Table~\ref{tab:lambda1_sweep} show the effect of the regularization parameter \(\lambda_1\). For the noise-free scenarios ((a) and (c)), the reconstruction error decreases monotonically as \(\lambda_1\) decreases down to \(10^{-9}\). This is explained by the fact that in the absence of random fluctuations, the stabilizer \(\Omega[f]\) penalizes only physically unjustified high-frequency oscillations, and its weakening allows the neural network to detail the spectrum more accurately. Here, the fixed value \(\lambda_2 = 0.01\) ensures stable compensation of the kernel's systematic error without risking parasitic distortions to the shape of the reconstructed spectrum.

On the other hand, for noisy data (scenarios (b) and (d)), the error saturates: upon passing from \(10^{-7}\) to \(10^{-9}\), the \(L_2\) error decreases insignificantly (from 12.72~\% to 12.42~\% in scenario (b)), and in scenario (d) it practically reaches a plateau (from 4.40~\% to 4.15~\%). Further reduction of \(\lambda_1\) no longer yields substantial improvement in reconstruction. This is because \(\lambda_2 = 0.01\) limits the freedom of the correction \(\Delta K_\phi\), and under strong noise the main source of error becomes precisely the noise in the right-hand side \(g_\delta(x)\) rather than the inaccuracy of the kernel. Thus, the chosen value \(\lambda_2 = 0.01\) provides a reasonable compromise: it allows the systematic error of the operator to be compensated without shifting the burden to the kernel to adjust to the random measurement noise.

We have shown that a further decrease of the regularization parameter below \(\lambda_1 = 10^{-8}-10^{-9}\) no longer leads to significant changes in $f(s)$. This is where the PINN-based approach differs from the classical regularization method \cite{Tikhonov_1977}, which was widely used earlier, for example, in works on determining the spectral energy composition of electron beams from vacuum and gas diodes. Thus, in \cite{Baksht_2010, Kozyrev_2015}, spectrum reconstruction was also performed by successively decreasing the parameter $\alpha$ (analogous to $\lambda_1$) while monitoring the approach of the computed right-hand side of equation \eqref{Fredholm} to the specified one using the maximum relative deviation. The stopping criterion was the appearance of unphysical distribution artifacts in $f(s)$, the most significant of which was the violation of spectral nonnegativity, i.e., $f(s)<0$. Some sets of attenuation curves were rejected using the classical method \cite{Tikhonov_1977}, even after multiple preliminary processing (densification) by interpolation. The solution to the regularization equation \eqref{Tihonov_eq} alternated in sign, even at a significant residual in the right-hand side (obtained for larger $\alpha$ values). In other cases, the change $\lambda_1 \xrightarrow{}0$ led to the achievement of $f(s)<0$ earlier than the accuracy was achieved by satisfying the condition on the residual of the right-hand side $\|\hat{A}f_\alpha - g_\delta\| < \delta$. In other words, in the classical method \cite{Tikhonov_1977}, the spectrum $f(s)$ lost its physical meaning before any nontrivial physical features, in the form of additional distribution maxima, appeared. If, however, one ignores the violation of the nonnegativity condition $f(s) \geq 0$ and continues solving the Tikhonov regularization equation \eqref{Tihonov_eq} by driving $\alpha$ to zero, then one can achieve satisfaction of the condition $\|\hat{A}f_\alpha - g_\delta\| < \delta$ (and even $\|\hat{A}f_\alpha - g_\delta\| \ll \delta$), although such a solution will no longer have any physical meaning.

The validation carried out on a model-synthesized data set confirms the effectiveness and physical validity of the proposed PINN-based approach on one of the examples with a known model spectrum \(f_{\text{model}}(s)\). The method reliably resolves the complex multi-peak structure of the spectrum under a specified fixed error of the integral operator kernel (the Tabata-Ito formula) \eqref{kernel} at the 10~\% level indicated in \cite{Tabata_1975}. 

\section{Reconstruction of the Electron Energy Spectra in Pulsed Vacuum Diode}

\begin{figure}[!htb]
\centering
\includegraphics[scale=.2]{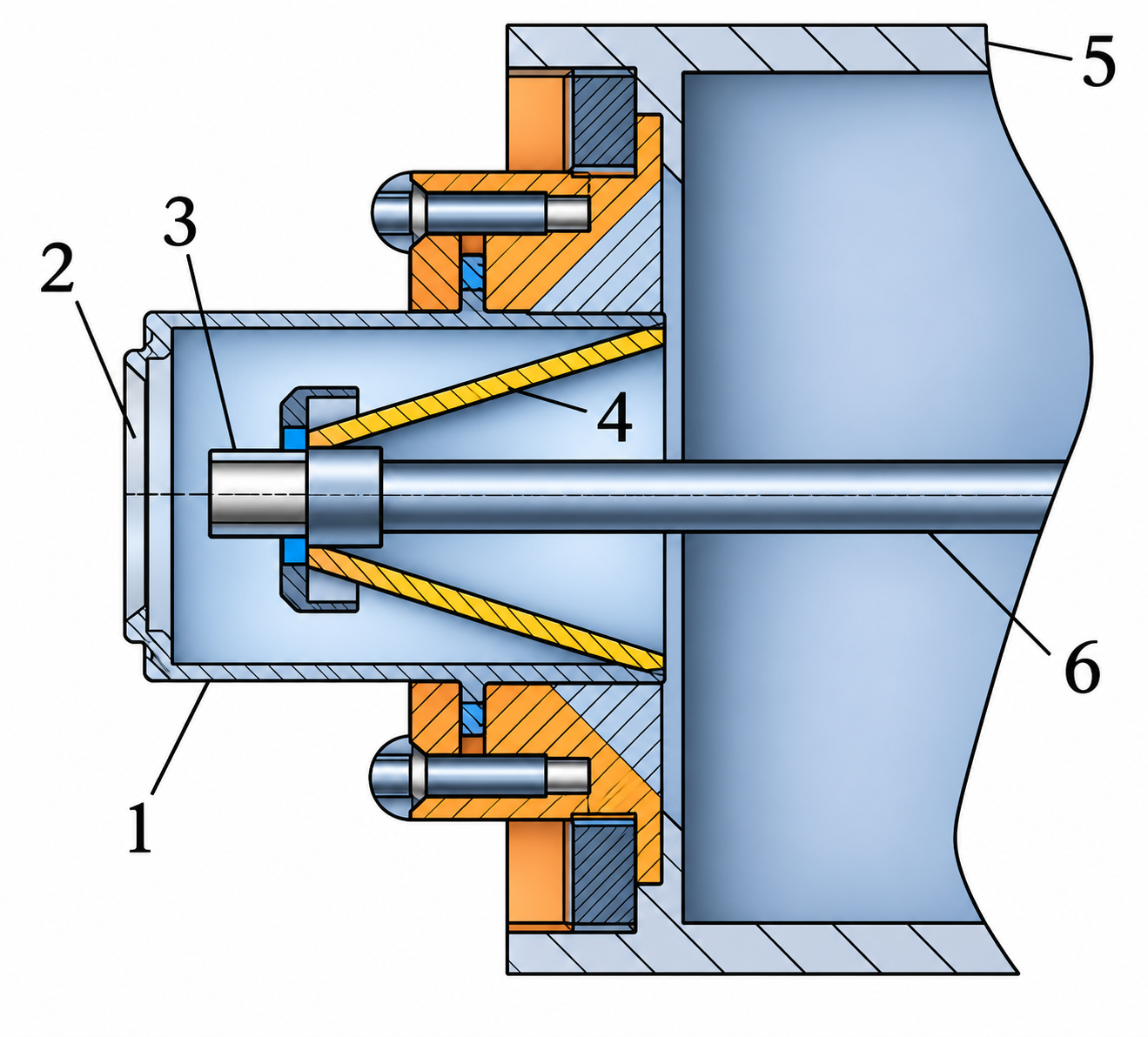}
\caption{\label{fig:IMAZ} Design of the output part of the RADAN-220 generator with a sealed-off vacuum tube IMA3-150E: 1 – electron tube housing; 2 – beryllium foil window 200~$\mu$m thick; 3 – tubular cathode 5~mm in diameter made of steel foil; 4 – insulator; 5 – part of the RADAN-220 generator housing; 6 – high-voltage output of the generator placed in transformer oil.}
\end{figure}

The successful validation of the PINN-based methodology for solving the full inverse problem for the Fredholm equation \eqref{Fredholm} on a large number of synthetic model data created a solid foundation for proceeding to the processing of real experimental attenuation curves obtained in nanosecond gas-filled and vacuum diodes \cite{Kozhevnikov_2026, Kozhevnikov_2026_preprint1, Kozhevnikov_2026_preprint2}. Below, we consider an example of applying the improved technique to indirectly measure the spectral energy characteristics of electrons at the output of a pulsed compact nanosecond accelerator. It consists of a nanosecond RADAN-220 generator \cite{Yalandin_2001} connected to a vacuum diode (a sealed-off vacuum electron beam tube), the IMA3-150E \cite{Baksht_2007}. The design of the output part of the RADAN-220 generator with the IMA3-150E tube at its left end is shown in Fig.~\ref{fig:IMAZ} 

The RADAN-220 generator produced negative-polarity voltage pulses with a full width at half maximum of 3~ns and a leading edge of 0.5~ns. The voltage amplitude across the IMA3-150E tube, which was installed at the end of the generator, reached 200~kV. The electron beam current was measured using a 20~mm-diameter collector connected to a cable with a wave impedance of 50~$\Omega$. The collector's temporal resolution was no worse than 80~ps. The oscillogram of the beam current pulse behind a beryllium foil 200~$\mu$m thick is shown below in Fig.~\ref{fig:current_pulse} 

\begin{figure}[!htb]
\centering
\includegraphics[scale=.45]{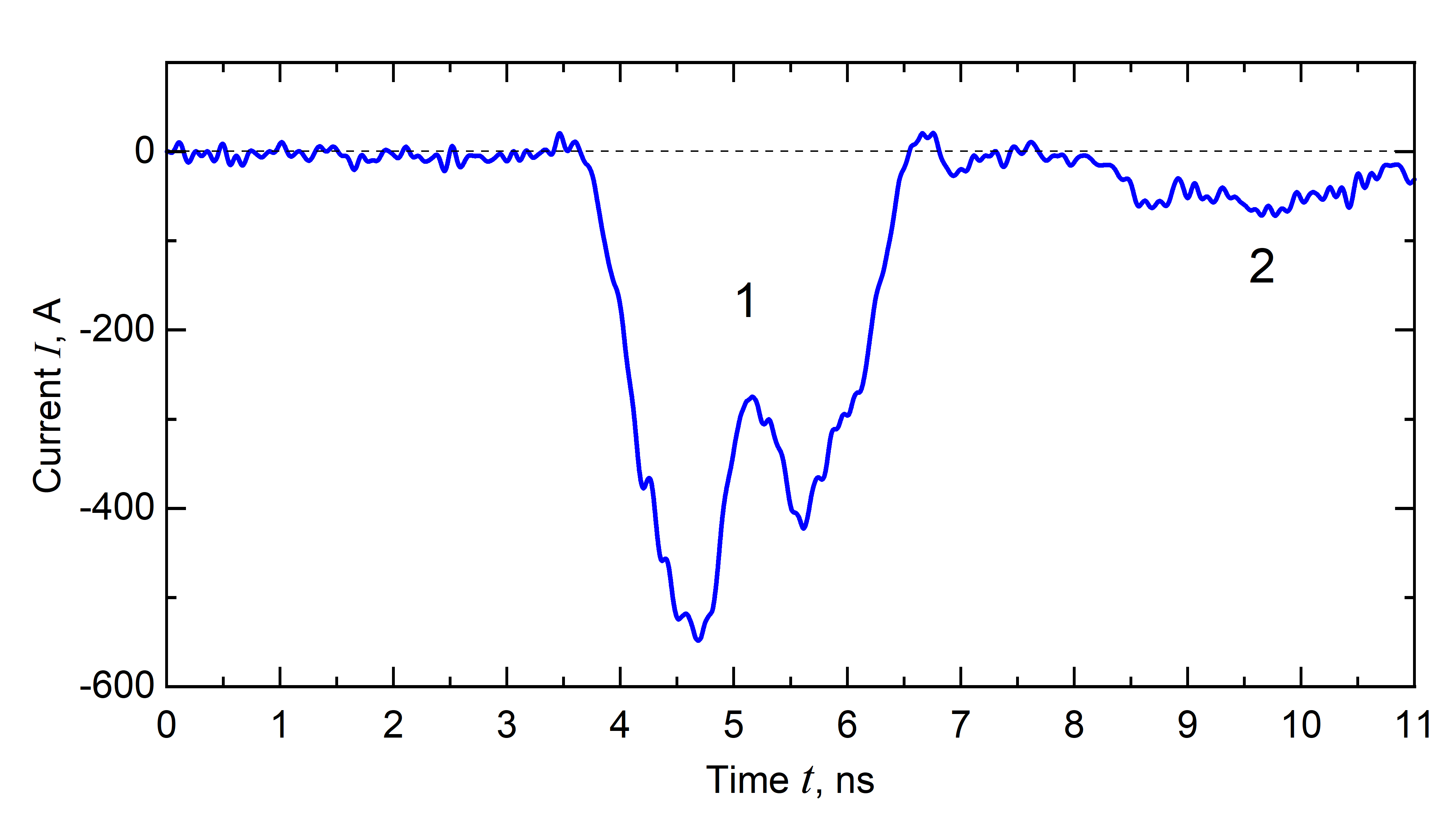}
\caption{\label{fig:current_pulse} Oscillogram of the electron beam current behind the anode of the sealed-off vacuum tube IMA3-150E connected to the nanosecond pulsed voltage generator RADAN\mbox{-}220 (200-$\mu$m beryllium foil). Here, two components are distinguished: main pulse (1) with two peaks and post-pulse (2), delayed by $\sim$1~ns and having a substantially lower amplitude.}
\end{figure}

\begin{figure}[!htb]
\centering
\includegraphics[scale=.25]{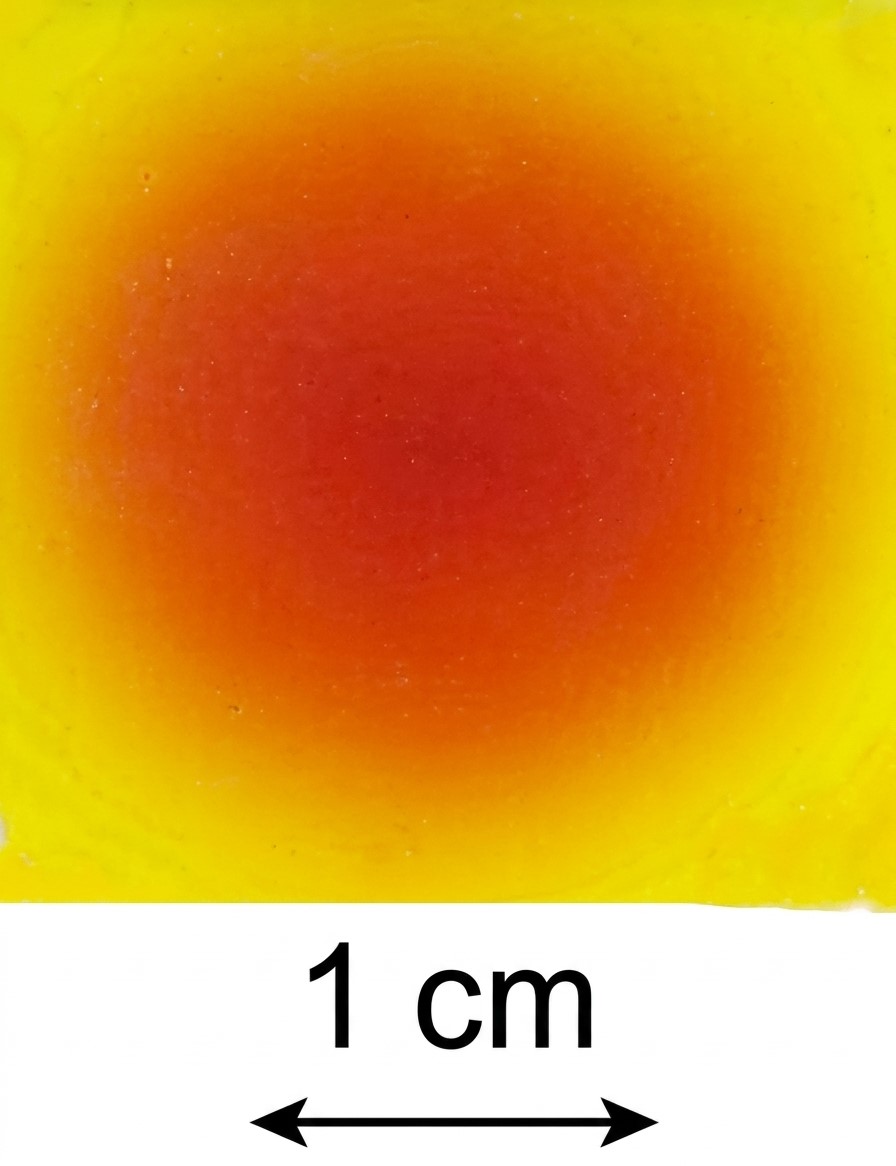}
\caption{\label{fig:autograf} The electron beam autograph obtained on a vinylpros film installed 1~cm from the foil over 20 cycles of generator activation. The original image was upscaled to 600~dpi for print clarity without altering its color distribution, intensity profile, or spatial geometry.}
\end{figure}

In the beam current oscillogram in Fig.~\ref{fig:current_pulse}, two pulses stand out: the first (1) has two peaks, and the second (2) is delayed by 1~ns and has a substantially smaller amplitude. The characteristic autograph of the beam current (Fig.~\ref{fig:autograf}) has a diameter of 2~cm. With a large number (more than twenty) of accelerator activations, the beam autograph is uniform in cross section. However, with a single activation, the autograph consists of three regions with higher current density against a less dense, uniform background with a diameter of 2~cm \cite{Baksht_2007}.

To construct the attenuation curves, the electron charge at the current collector was used, calculated from the measured amplitude and duration of the beam current behind the anode, composed of beryllium and aluminum foils. To record the current and voltage pulses, a Tektronix~TDS6604 real-time oscilloscope (6~GHz and 20~GS/s) was used. The signals were fed to the oscilloscope via an RG58-A/U high-frequency cable (Radiolab) 1~m long with N-type (Suhner~11~N-50-3-28/133~NE) and SMA (Radiall~R125.075.000) connectors. The beam attenuation data in aluminum foils of various thicknesses in the range of 150–500~$\mu$m were measured both for the first part (1) of the pulse and for the total pulse (Fig.~\ref{fig:current_pulse}). Thus, the PINN-based algorithm had two attenuation curves at its disposal, each consisting of $N_{exp} = 24$ points measured with a relative error of no more than 10~\%. The fixed relative accuracy of the kernel in equation \eqref{Fredholm} was also specified to be within 10~\%, which represents the root-mean-square error of the Tabata-Ito formula \eqref{kernel} for the attenuation of an electron beam in aluminum foils \cite{Tabata_1975}. With a fixed value of the constant $\lambda_2 = 0.01$, in each cycle of neural network training, the parameter $\lambda_1$ was varied so that the residual of the right-hand side \eqref{g_model} computed from the reconstructed curve $f(s)$ fell within the confidence interval of the measured experimental values. For additional verification, training was carried out for several quadrature nodes varied over a wide range, $N_q = 1000..10000$.

\begin{figure}[!htb]
\centering
\includegraphics[scale=.45]{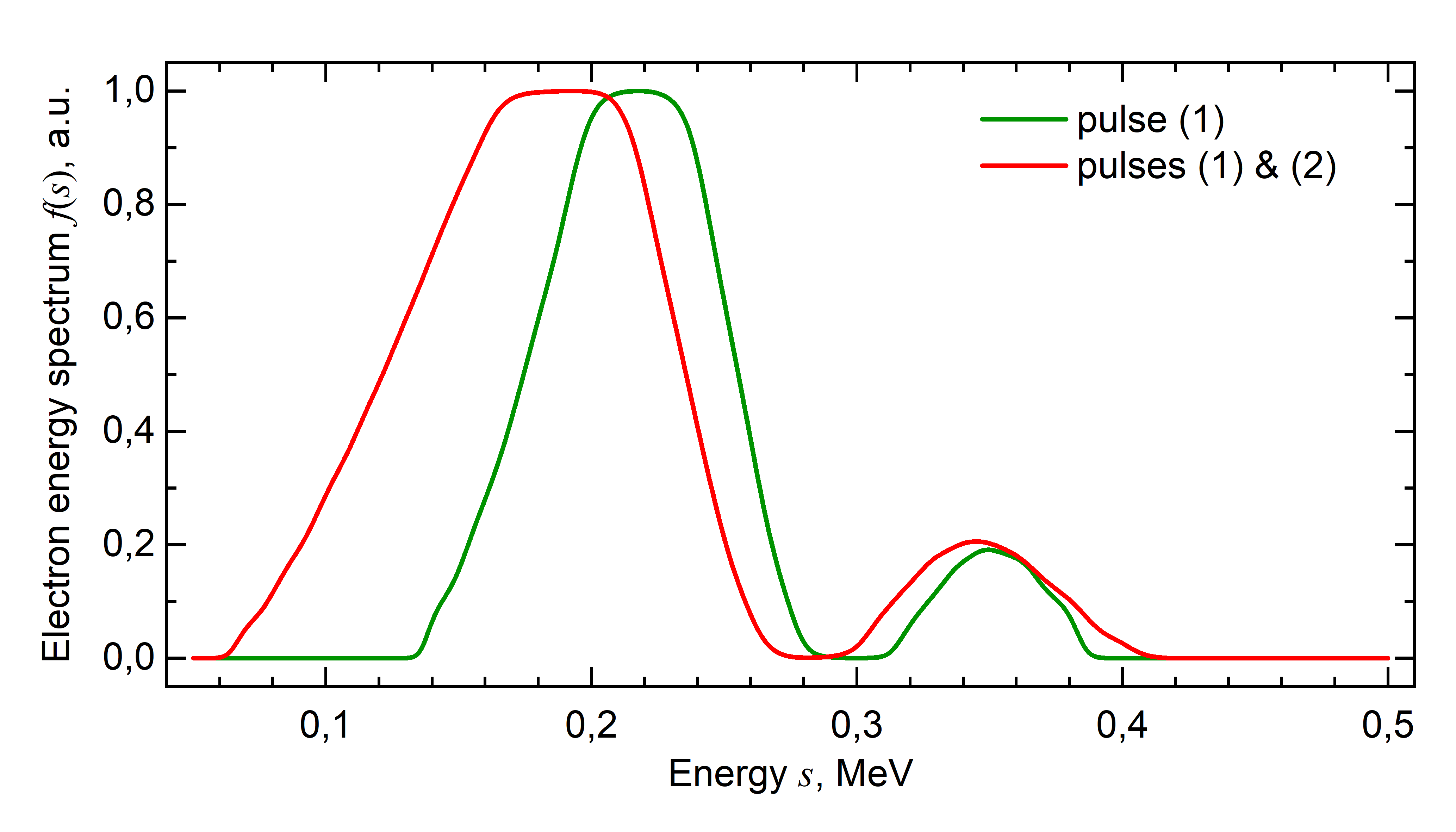}
\caption{\label{fig:IMAZ_spectra} Spectral energy distributions corresponding to the main pulse (1) alone (green line) and to the full oscillogram (1)-(2) (red line) of the beam current (Fig.~\ref{fig:current_pulse}).}
\end{figure}

Fig.~\ref{fig:IMAZ_spectra} presents the results of the PINN-based method for reconstructing the energy spectra of electrons in the installation shown in Fig.~\ref{fig:IMAZ}. It also shows the experimental attenuation curves, their confidence intervals, and the right-hand sides of the Fredholm equation obtained by integrating the reconstructed spectra $f(s)$ according to formula \eqref{g_model}. The electron beam spectra in the energy range of 50–500~keV consist of two clearly distinguishable groups. The main group of electrons is represented in Fig.~\ref{fig:IMAZ_spectra} by the main maximum (the curves are normalized to the maximum value). For the two-pulse structure (Fig.~\ref{fig:current_pulse}), the main spectral maximum at 192~keV (red curve in Fig.~\ref{fig:IMAZ_spectra}) is located to the left of the single-pulse spectrum maximum at 218~keV (green curve, same figure). The half-width of the main spectral maximum of the red curve is 115~keV, which is greater than the half-width of the main maximum for the pulse part (1) alone — 81~keV. The positions of both main maxima lie in the voltage region at RADAN-220 generator output; i.e., most beam electrons have energies $s \sim qU_0$, with $U_0 \sim 200$~kV. 

The total electron pulse, consisting of the main part (1) in Fig.~\ref{fig:current_pulse} and the post-pulse (2), has an expectedly larger spectral width of the main electron group. It extends into the low-energy region. The second spectral energy group in Fig.~\ref{fig:IMAZ_spectra} is the group of electrons with anomalously high energy (anomalous electrons). In both spectra, the second maximum lies at $\sim 350$~keV and has a half-width of about 60~keV; thus, the characteristics of the anomalous electrons are similar in both cases and, taking into account the experimental uncertainties, can be considered indistinguishable. In accordance with the rigorous kinetic theory of the formation of electrons with anomalously high energies in vacuum diodes \cite{Kozhevnikov_2022}, electrons with energies $s > qU_0$ appear only at the front of the anode voltage pulse, the duration of which is comparable to the transit time of electrons in the given gap. This regularity manifests itself equally in vacuum diodes with thermionic emission (electron) \cite{Kozhevnikov_2022} and in explosive-emission diodes, where a dense quasi-neutral plasma forms at the cathode \cite{Kozhevnikov_2023_1, Kozhevnikov_2023_2}. The physics of this process is that sharp leading edges of the anode voltage excite space-charge relaxation oscillations. The shorter the front duration, the higher the initial amplitude of the anomalous electrons' energy can be. In the idealized case, when the voltage is applied instantaneously (the front duration is zero), the theoretically predicted excess $s > qU_0$ can be up to 3 times \cite{Kadish_1985}. Naturally, under real experimental conditions, the leading-edge durations are always nonzero, and here the more rigorous kinetic theory \cite{Kozhevnikov_2022} for supply-voltage pulses with subnanosecond leading edges for planar gaps of 1~cm gives an average estimate of 1.3–1.5 times the excess amplitude energy. The post-pulse of the collector current (2) in Fig.~\ref{fig:current_pulse} is represented by low-energy electrons and therefore does not affect the position and half-width of the high-energy maximum of the curves in Fig.~\ref{fig:IMAZ_spectra}.

\begin{figure}[!htb]
\centering
\includegraphics[scale=.45]{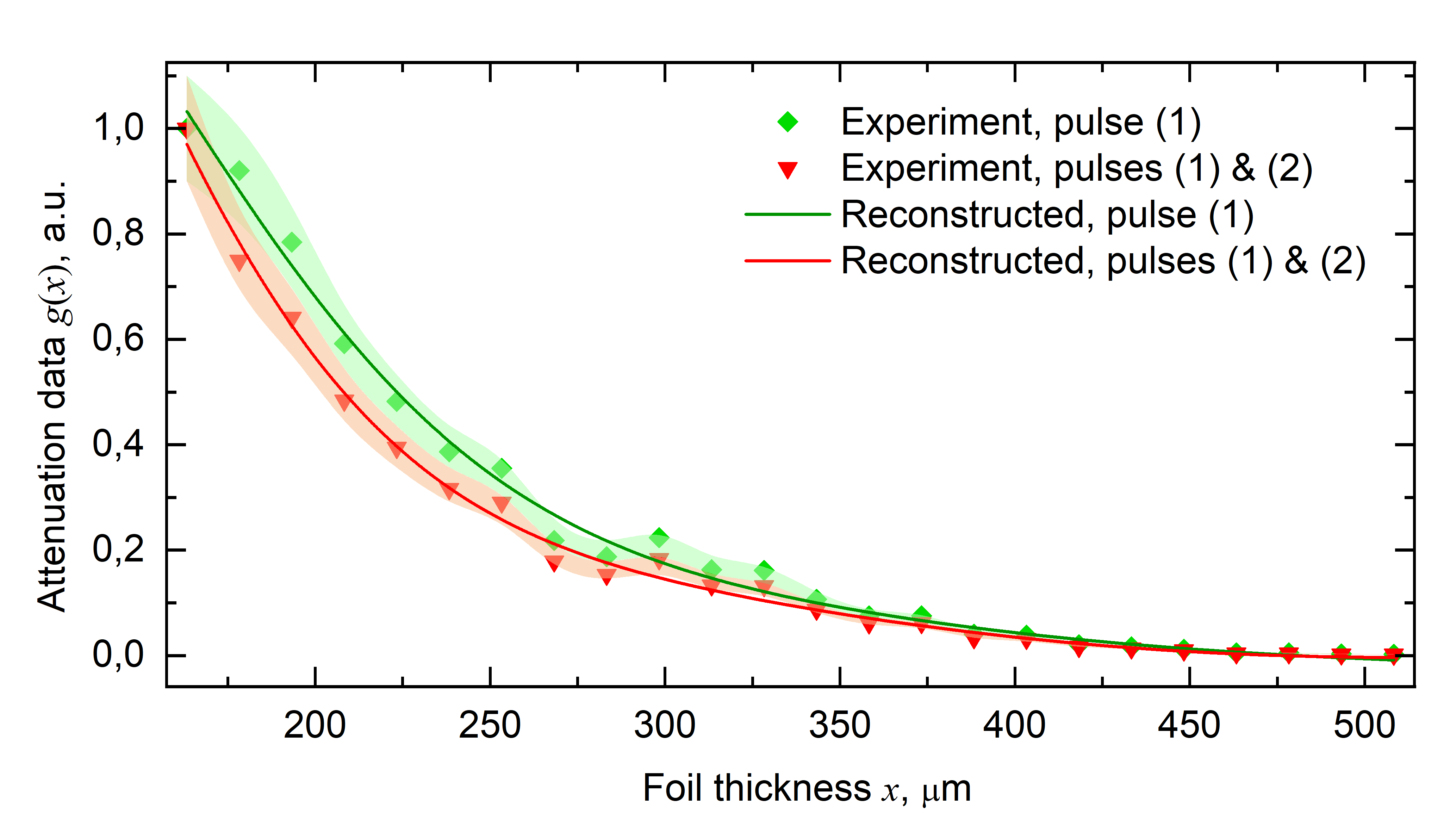}
\caption{\label{fig:IMAZ_attenuation} Experimental attenuation curves (discrete values) for the oscillogram without the post-pulse (2) (green) and for the full oscillogram (red) of the beam current (Fig.~\ref{fig:current_pulse}). The confidence intervals for both attenuation curves are shown as solid-color regions matching the color of the experimental data points. Also, continuous curves of the same colors show the attenuation curves calculated from the reconstructed spectra (Fig.~\ref{fig:IMAZ_spectra}) according to formula \eqref{g_model}.}
\end{figure}

\section{Conclusions and Summary}
This work presents a new mesh-free computational method for solving the ill-posed problem for the Fredholm equation of the first kind \eqref{Fredholm}, based on physics-informed neural networks with adaptive kernel correction. The proposed architecture includes two jointly trained deep neural networks. It allows the ill-posed problem of reconstructing (regularizing) the solution $f(s)$ to be solved in the most complete formulation, without preliminary preparation of input datasets through interpolation and approximation. In this context, the proposed approach is of practical value for processing experimental results related to solving the inverse problem for equation \eqref{Fredholm}.

The validation example on synthetic multi-peak spectra shown in the work convincingly demonstrates that the method is stable under 10~\% random noise and significant sparsification of the initial data, and that using an adaptive kernel (DeltaKNet) reduces the root-mean-square reconstruction error. The conditions modeled in the example are analogous to those in problems of determining the energy composition of electron beams in installations with nanosecond and subnanosecond discharge durations. Applying the method to real experimental attenuation curves to obtain the electron energy spectrum of a nanosecond vacuum diode confirmed its operability under the indicated conditions; in particular, it revealed a nontrivial structure in the electron spectrum with anomalously high electron energies. 

The proposed algorithm applies to a broad class of inverse problems of mathematical physics of the form \eqref{Fredholm}, in which all input parameters are known only with some error. The demand for such a tool is driven by the growing number of experimental situations — in beam and plasma physics, spectroscopy, tomography, and remote sensing — where the measured signal is an integral response and the sought-for distribution cannot be obtained by direct inversion. Several directions for further development appear particularly promising. First, the method can be extended to two- and three-dimensional Fredholm and Radon-type operators by combining separable or low-rank kernel representations with tensor-product or sparse-grid quadrature, domain decomposition, and GPU-parallel training; this would make the framework applicable to tomographic reconstruction and spatially resolved spectroscopy. Second, a systematic uncertainty-quantification layer is needed: Bayesian PINNs, deep ensembles, or calibrated Monte Carlo dropout could provide credible intervals for the reconstructed spectrum and separate the contributions of data noise, kernel-model error, quadrature error, and network approximation error. Third, the adaptive kernel correction can be further constrained by physical invariants (nonnegativity, monotonicity, known asymptotics) and combined with bilevel or generalized-discrepancy procedures to automatically select $\lambda_1$ and $\lambda_2$, so the regularization parameters are calibrated to the measured error level rather than tuned manually. Fourth, the computational cost can be reduced by amortized inference. Once trained, Fourier neural operators (FNOs) or DeepONet-type surrogates could map attenuation curves directly to spectra, enabling rapid re-analysis and near-real-time reconstruction on edge hardware. Fifth, integrating with other diagnostics (voltage and current waveforms, X-ray signals, multi-angle attenuation data) and with digital-twin models of the diode would enable multimodal fusion and optimal experimental design, including adaptive choice of foil thicknesses. Finally, the theoretical agenda includes convergence and stability guarantees for the joint spectrum–kernel learning problem and validation on benchmark datasets with known ground truth. Such developments would turn the proposed approach from a proof of concept into a general uncertainty-aware tool for ill-posed indirect measurements beyond charged-particle spectroscopy.

\begin{acknowledgments}
This work was conducted without any external funding, including state assignments. The authors express their gratitude to colleagues and like-minded people for fruitful discussions, but the research itself was carried out solely on the authors' initiative. Future funding to continue this line of research would be highly appreciated.    
\end{acknowledgments}

\bibliography{KKK}

\end{document}